\documentclass[aps,twocolumn,prl,showpacs,preprintnumbers,
superscriptaddress]{revtex4-1}
\usepackage{rotating}
\usepackage[mathscr]{eucal}

\usepackage{amsmath,amssymb,bm,comment}
\usepackage{graphicx}
\usepackage{epsfig}
\usepackage{epstopdf}
\usepackage{hyperref}
\usepackage{longtable}

\def\k{{\mathbf k}}
\def\J{{\cal J}}
\def\T{{\cal T}}

\begin{document}

\title{Triangle Anomalies, Thermodynamics, and Hydrodynamics}
\author{Kristan Jensen}
\affiliation{Department of Physics, University of Victoria,
Victoria, BC V8W 3P6, Canada}
\date{\today}
\begin{abstract}
We consider 3+1-dimensional fluids with $U(1)^3$ anomalies. We use Ward identities to constrain low-momentum Euclidean correlation functions and obtain differential equations that relate two and three-point functions. The solution to those equations yields, among other things, the chiral magnetic conductivity. We then compute zero-frequency functions in hydrodynamics and show that the consistency of the hydrodynamic theory also fixes the anomaly-induced conductivities.
\end{abstract}
\maketitle 

\emph{Introduction.}---%
The chiral magnetic effect (CME)~\cite{Kharzeev:2007tn,Fukushima:2008xe,Kharzeev:2009pj} and chiral vortical effect (CVE)~\cite{Son:2009tf,Kharzeev:2010gr} represent remarkable implications of anomalies in quantum field theory for macroscopic transport. In a fluid with a $U(1)^3$ anomaly, there will be currents directed along a magnetic field $B^{\mu}$ or the vorticity $w^{\mu}$. The former is the CME and the latter the CVE. Both effects probe the violation of parity and charge conjugation and so may be measured by studying spatial and charge asymmetries~\cite{Kharzeev:2010gr} in off-axis heavy ion collisions at RHIC or the LHC. 

It has also been shown that the hydrodynamic description of a relativistic fluid with anomalies must be modified~\cite{Son:2009tf} from its textbook treatment~\cite{LL6}. There are additional transport coefficients, describing the response of the currents to magnetic fields and vorticity. These coefficients are fixed by demanding a local version of the second law of thermodynamics, namely the existence of an entropy current whose divergence is positive semi-definite~\cite{LL6}. Despite the absence of a clear understanding of this constraint in field theory, the result for the anomaly-induced transport matches results at weak~\cite{Fukushima:2008xe} and strong coupling~\cite{Banerjee:2008th,Erdmenger:2008rm}. This yields a hydrodynamic description of the CME and CVE.

We endeavor to reproduce the constraints on anomaly-induced transport without recourse to the entropy current. We will do so by employing properties of equilibrium quantum field theory. In particular, we will study theories which have finite static correlation length, $\lambda$. This has the practical implication that real-space, zero-frequency correlation functions fall off exponentially at long distance. The Fourier transformed functions are then analytic at zero frequency and small momenta $k\lambda\ll 1$. Together with some Ward identites we relate the $U(1)^3$ anomaly coefficient to transport. 

The resulting calculation makes a few points clear. First, only the small $B^{\mu}$ and small $w^{\mu}$ parts of the CME and CVE are fixed by the anomaly. Second, anomaly-constrained transport follows from a covariant and gauge-invariant description of thermodynamics. Finally, we see that, among other things, the entropy current enforces properties of the equilibrium theory (see~\eqref{E:recip} and~\eqref{E:TmuP}) that are not manifest in the hydrodynamic description.

\emph{Note:} We explore a number of questions related to equilibrium thermodynamics and the role of the entropy current in a companion paper~\cite{Jensen:2012jh}. While this work was in progress we were also made aware of~\cite{Banerjee:2012iz} which has overlap with the content of this Letter.

\emph{Correlation functions at $T,\mu\neq 0$.}---%
In real-time finite temperature field theory there are different definitions for correlation functions. We employ the closed-time-path (CTP) formalism, in which time is extended to a closed contour which extends from $t_1 :(-\infty,\infty)$, and then doubles back as $t_2:(\infty,-\infty)$. See~\cite{Wang:1998wg} for a review. The only ingredient we need is that there is a CTP generating functional $W_{CTP}$ which depends on two sets of sources: $J_r = (J_1+J_2)/2$, $J_a = J_1-J_2$, where the $J_i$ are independent sources introduced on each segment of the time contour. The fully retarded functions, which are those computed in hydrodynamics~\cite{Moore:2010bu}, are obtained by varying $W_{CTP}$ with respect to the $r$ sources and one $a$ source. The $r$ source couples to $a$-type operators, whereas the $a$ source couples to $r$-type operators. Insertions of $U(1)$ currents, labeled by the index $A$, and stress tensor densities come from varying $W_{CTP}$ with respect to background fields as
\begin{equation}
\langle \J^{\mu,A}_{r/a}(x)\rangle = \frac{\delta W_{CTP}}{\delta A^{A}_{\mu,a/r}(x)}, \,\,\,\, \langle \T^{\mu\nu}_{r/a}(x)\rangle = \frac{2\,\delta W_{CTP}}{\delta g_{\mu\nu,a/r}(x)}.
\end{equation}
The currents and stress tensor are related to the densities by $\J^{\mu,A}=\sqrt{-g}J^{\mu,A}, \T^{\mu\nu}=\sqrt{-g}T^{\mu\nu}$. In this work we consider $ra$ and $raa$ functions in momentum space. We notate these as
\begin{align}
G_R^{I,J}(q)& = \langle \mathcal{O}^I_r(q)\mathcal{O}^J_a(-q)\rangle, \\
G_R^{I,J,K}(q_1,q_2)& = \langle \mathcal{O}^I_r(q_1)\mathcal{O}^J_a(q_2)\mathcal{O}^K_a(-q_1-q_2)\rangle,
\end{align}
where coordinate space functions are related to their momentum space cousins by
\begin{equation}
f(x_1,..,x_n) = \int d^dq_1..d^dq_n e^{i (q_1\cdot x_1 + .. q_n\cdot x_n)}f(q_1,..,q_n).
\end{equation}
We further specialize to zero frequency $ra..a$ functions, i.e. we take $q_i=(0,\k_i)$. These (and all other CTP functions) are proportional to the corresponding Euclidean time-ordered functions obtained by variation of the Euclidean generating functional $W_E$~\cite{Evans:1991ky,*Evans:1995ug}. Because of this we henceforth neglect the $r$ and $a$ indices. Additionally, since bosonic derivatives of $W_E$ commute, the CTP functions of bosonic fields satisfy
\begin{align}
\label{E:recip}
\langle .. \mathcal{O}^I(\k_1).. \mathcal{O}^J(\k_2) .. \rangle = \langle .. \mathcal{O}^J(\k_2) .. \mathcal{O}^I(\k_1).. \rangle.
\end{align}

The diffeomorphism invariance and anomalous variation of $W_{CTP}$ leads to the Ward identities~\cite{Herzog:2009xv},
\begin{align}
\label{E:Jcons}
\nabla_{\mu} \J^{\mu,A} &= -\frac{1}{24}C^{ABC}\epsilon^{\mu\nu\rho\sigma}F_{\mu\nu}^BF_{\rho\sigma}^C, \\
\label{E:Tcons}
\nabla_{\mu} \T^{\mu\nu} & = F^{\nu\rho,A}\left(\J_{\rho}^A-\frac{C^{ABC}}{6}\epsilon_{\rho}^{\phantom{\rho}\sigma\alpha\beta}A_{\sigma}^BF_{\alpha\beta}^C\right),
\end{align}
where $C^{ABC}$ is the symmetric anomaly coefficient and $\epsilon^{0123}=1$ . \eqref{E:Jcons} and~\eqref{E:Tcons} encode local gauge and diffeomorphism covariance. There are additional Ward identities for the Euclidean theory at nonzero temperature due to the global topology of $\mathbb{R}^3\times\mathbb{S}^1$~\cite{Jensen:2011xb}. Constant $A_0$ and $h_{00}$ may be gauged away at the price of redefining the temperature and chemical potentials. These are shifted as
\begin{align}
\label{E:TmuP}
T' = \frac{T}{\sqrt{-g_{00}}}, \qquad \mu'^{A} = \frac{A_0^A}{\sqrt{-g_{00}}}.
\end{align}
The $\mu^A$ depend on $g_{00}$ because they are defined through the Wilson line of $A^A$ around the time circle.

At small momenta $|\k_i| \lambda\ll 1$, the correlation functions of the current and stress tensor are heavily constrained. The two-point functions with parity-violating terms to $O(k)$ may be parametrized as
\begin{align}
\begin{split}
\label{E:G2pt}
G_R^{iA,jB}(\k) &=-i \epsilon^{ijk}k_k \left(\sigma_1^{AB}-\frac{C^{ABC}A_0^C}{3}\right), \\
G_R^{iA,0j}(\k) & =-i \epsilon^{ijk}k_k \sigma_2^A, \\
G_R^{0i,0j}(\k) & = \alpha_3 \delta^{ij} - i \epsilon^{ijk}k_k \sigma_3,
\end{split}
\end{align}
for some functions $\sigma_m$. By~\eqref{E:recip} $\sigma_1^{AB}=\sigma_1^{BA}$.  The second term in $G_{R}^{iA,jB}$ comes from the Bardeen-Zumino polynomial, which encodes the anomalous dependence of the $\J^{\mu,A}$ on gauge fields~\cite{Bardeen:1984pm}. The three-point functions with parity-violating terms to $O(k)$ are similarly constrained. 
Demanding that the three-point functions are consistent with~\eqref{E:recip} we find that they take the form
\begin{align}
\begin{split}
\label{E:G3pt}
G_R^{iA,jB,0C}(\k_1,\k_2) & = {-} i \epsilon^{ijk}((k_1)_k \Sigma^{0,ABC}_{1}{-}(k_2)_k\Sigma_{1}^{0,BAC}) , \\
G_R^{iA,jB,00}(\k_1,\k_2) & ={-} i \epsilon^{ijk}((k_1)_k \Sigma^{00,AB}_{1}{-}(k_2)_k\Sigma_{1}^{00,BA}) , \\
G_R^{iA,0j,0B}(\k_1,\k_2) &={-} i \epsilon^{ijk} ((k_1)_k\Sigma^{0,AB}_{2,1} {+} (k_2)_k \Sigma^{0,AB}_{2,2}), \\
G_R^{iA,0j,00}(\k_1,\k_2) & = {-} i \epsilon^{ijk} ((k_1)_k\Sigma^{00,A}_{2,1} {+} (k_2)_k \Sigma^{00,A}_{2,2}), \\
G_R^{0i,0j,0A}(\k_1,\k_2) & = \alpha_3^{0,A}\delta^{ij}  {-} i \epsilon^{ijk}(k_{1}-k_2)_k \Sigma^{0,A}_3, \\
G_R^{0i,0j,00}(\k_1,\k_2) & = \alpha_3^{00}\delta^{ij} {-} i \epsilon^{ijk}(k_1{-}k_2)_k \Sigma^{00}_3.
\end{split}
\end{align}

Imposing~\eqref{E:Jcons} fixes half of the $\Sigma$'s. For example, we may compute $i(k_1)_iG_R^{iA,jB,0C}$ directly from~\eqref{E:G3pt} or by variation of~\eqref{E:Jcons}. Setting the two equal gives
\begin{align}
i (k_1)_i G_R^{iA,jB,0C}(\k_1,\k_2) &=\epsilon^{jkl}(k_1)_k(k_2)_l \Sigma_{1}^{0,ABC} \\
\nonumber
&=\frac{\delta^2\langle \partial_{\mu} \J^{a\mu,A}(\k_1)\rangle}{\delta A_j^B({-}\k_2)\delta A_0^C(\k_1{+}\k_2)} \\
\nonumber
&={-}\frac{1}{24} \frac{\delta^2 C^{ABC}\epsilon^{\mu\nu\rho\sigma}(F^B_{\mu\nu}F^C_{\rho\sigma})(\k_1)}{\delta A_j^B({-}\k_2)\delta A_0^c(\k_1{+}\k_2)}\\
\nonumber & = \epsilon^{jkl}(k_1)_k(k_2)_l \frac{C^{ABC}}{3}.
\end{align}
By applying the same method to any $G_R$ with a $\J^{iA}$ insertion we thereby find
\begin{align}
\label{E:Sigma1}
\Sigma_{1}^{0,ABC} = \frac{C^{ABC}}{3},\,\, \Sigma_{1}^{00,AB} =\Sigma_{2,2}^{0,AB}=\Sigma_{2,2}^{00,A}=0.
\end{align}
Imposing~\eqref{E:Tcons} fixes the remaining $\Sigma$'s. For example, 
\begin{align}
i(k_2)_jG_R^{iA,0j,0B}(\k_1,\k_2) &= - \epsilon^{ikl}(k_1)_k(k_2)_l \Sigma_{2,1}^{0,AB} \\
\nonumber &= \frac{\delta^2\langle\partial_{\mu} \T^{\mu 0}(\k_2)\rangle}{\delta A_i^A({-}\k_1)\delta A_0^B(\k_1{+}\k_2)} \\
\nonumber & = i (k_1{+}k_2)_k G_R^{kB,iA}({-}\k_1),
\end{align}
which by~\eqref{E:G2pt} and~\eqref{E:recip} gives
\begin{equation}
\label{E:Sigma2}
\Sigma_{2,1}^{0,AB}=\sigma_1^{AB}.
\end{equation}
Applying this method to the other three-point functions in~\eqref{E:G3pt} yields the remaining $\Sigma$'s
\begin{equation}
\label{E:Sigma3}
 \Sigma_{2,1}^{00,A} = 2\sigma_2^A, \,\, \Sigma_3^{0,A} = 2 \sigma_2^A, \,\, \Sigma_3^{00}=4\sigma_3.
\end{equation}

By the discussion around~\eqref{E:TmuP} we may evaluate the two-point functions~\eqref{E:G2pt} in a background with constant $A_0^A$ and $g_{00}$. To $O(k)$ we find
\begin{align}
\begin{split}
\label{E:G2ptS}
G_{R,S}^{iA,jB}(\k) &= - i \epsilon^{ijk}k_k\left(\sqrt{-g_{00}}\sigma_1'- \frac{C^{ABC}A_0^C}{3}\right), \\
G_{R,S}^{iA,0j}(\k) & = - i \epsilon^{ijk}k_k \sigma_2'^A, \\
G_{R,S}^{0i,0j}(\k) &= -\frac{\alpha_3'\delta^{ij}}{g_{00}}-i\epsilon^{ijk}k_k\frac{\sigma_3'}{\sqrt{-g_{00}}},
\end{split}
\end{align}
where the prime indicates that a quantity is evaluated at temperature $T'$ and chemical potentials $\mu'^A$~\eqref{E:TmuP}, and the subscript $S$ that the correlator is evaluated in the presence of background fields.

Differentiating~\eqref{E:G2ptS} with respect to $A_0^A$ and $g_{00}$ leads to three-point functions with zero momentum insertions of $\J^{0,A}$ and $\T^{00}$. Comparing these functions with the three-point functions~\eqref{E:G3pt} and using~(\ref{E:Sigma1}), (\ref{E:Sigma2}), and~(\ref{E:Sigma3}), the two agree only if the six equations
\begin{align}
\label{E:eqsSigma}
\frac{\partial \sigma_1^{AB}}{\partial\mu^C} &= C^{ABC}, \,\,\,\, \frac{\partial\sigma_2^A}{\partial\mu^B}  = \sigma_1^{AB}, \,\,\,\,\frac{\partial\sigma_3}{\partial\mu^A}= 2\sigma_2^A, \\
\label{E:eqsE}
E_m & = T \frac{\partial \sigma_m}{\partial T} + \mu^A\frac{\partial \sigma_m}{\partial \mu^A} - m \sigma_m = 0,
\end{align}
are satisfied. These equations uniquely fix the $\sigma_m$ up to integration constants. We have
\begin{align}
\begin{split}
\label{E:sigmas}
\sigma_1^{AB} &= C^{ABC}\mu^C + f_1^{AB}T,\qquad f_1^{AB}=f_1^{BA},\\
\sigma_2^A & = \frac{1}{2}C^{ABC}\mu^B\mu^C + f_1^{AB}\mu^B T + f_2^A T^2, \\
\sigma_3 & = \frac{1}{3}C^{ABC}\mu^A\mu^B\mu^C + f_1^{AB}\mu^A\mu^B T \\ & \phantom{=\,\,}+ 2 f_2^A \mu^A T^2 + f_3 T^3.
\end{split}
\end{align}

We may also consider the behavior of the integration constants $f_m$ under CPT. By a hydrodynamic argument employed in~\cite{Bhattacharya:2011tra}, we find that the $f_1^{AB}$ and $f_3$ are CPT-violating, while the $f_2^A$ are CPT-preserving. We can also directly establish this result by studying the transformation of the two-point functions~\eqref{E:G2pt} under CPT.


\emph{Hydrodynamics with sources.}---%
It is instructive to reproduce~\eqref{E:sigmas} from hydrodynamics. In this section we begin with equilibria at constant $T$, $\mu^A$, and vanishing sources. We take the fluid rest frame to be $u^{\mu}= v^{\mu}/\sqrt{-v^2}$, with $v^{\mu}$ a constant timelike vector. In these states the stress tensor and current are
\begin{equation}
\label{E:TJequil}
\langle T^{\mu\nu}\rangle = \epsilon u^{\mu}u^{\nu}+P\Delta^{\mu\nu}, \qquad \langle J^{\mu,A}\rangle = \rho^A u^{\mu},
\end{equation}
with $P$ and $\epsilon$ the pressure and energy density, obeying
\begin{equation}
dP = s dT + \rho^A d\mu^A, \qquad \epsilon +P = s T + \mu^A \rho^A,
\end{equation}
and $\Delta^{\mu\nu}=g^{\mu\nu}+u^{\mu}u^{\nu}$ is a projector satisfying $\Delta^{\mu\nu} u_{\nu} = 0, \Delta^2=\Delta$. In hydrodynamics we study long-wavelength fluctuations around equilibrium states. Those fluctuations may be described by promoting $T,$ $\mu^A$, and the $u^{\mu}$ to spacetime fields (the hydrodynamic variables) and expanding the stress tensor and current in gradients thereof -- this is the derivative expansion!\cite{Bhattacharyya:2008jc}. We also turn on slowly varying background gauge fields $A^A$ and metric perturbations $g=\eta+h$. We take the sources to be $O(1)$, so that the field strengths $F$ and connection coefficients $\Gamma$ are $O(\partial)$ in the derivative expansion. This is the scaling required to study the response of a fluid to sources.

In more general states we have the decomposition
\begin{align}
\nonumber
\langle T^{\mu\nu}\rangle &= \mathcal{E} u^{\mu} u^{\nu} + \mathcal{P} \Delta^{\mu\nu} + q^{\mu}u^{\nu}+q^{\nu}u^{\mu} + \tau^{\mu\nu}, \\
\label{E:TJdecomp}
\langle J^{\mu,A}\rangle & = \mathcal{N}^Au^{\mu}  + \nu^{\mu,A} +\frac{C^{ABC}}{6\sqrt{-g}} \epsilon^{\mu\nu\rho\sigma}A^B_{\nu}F^C_{\rho\sigma},\\
\nonumber \qquad u^{\mu} q_{\mu} &= u^{\mu} \nu^A_{\mu} = u^{\mu}\tau_{\mu\nu} = \Delta^{\mu\nu}\tau_{\mu\nu}=0.
\end{align}
\eqref{E:TJdecomp} has some redundancy. Fixing $\langle T\rangle$ and $\langle J^A\rangle$, we may redefine the hydrodynamic variables by $O(\partial)$ quantities to impose a choice of hydrodynamic frame. In the rest of this work, we perturb the equilibrium state~\eqref{E:TJequil} by sources which are static in the fluid rest frame, supposing that the perturbed fluid remains in a time-independent equilibrium. 
In such a state the one-point functions~\eqref{E:TJdecomp} and hydrodynamic variables will be local functions of sources -- the non-locality is on order of the size of the static screening lengths, which vanish in the hydrodynamic approximation. We then perform a change of frame such that $T$, $\mu^A$, and $v^{\mu}$ take on their form for equilibria with constant background fields,
\begin{equation}
\label{E:TmuEquil}
T = \frac{T_{\rm eq}}{\sqrt{-v^2}}, \qquad \mu^A = u^{\mu}A_{\mu}^A,\qquad  v^{\mu}=v^{\mu}_{\rm eq},
\end{equation}
where $T_{\rm eq}$ and $v^{\mu}_{\rm eq}$ are the temperature and velocity field of the source-free equilibrium state. This is the covariant version of~\eqref{E:TmuP}.

It remains to express $\mathcal{E}$, $\mathcal{P}$, $\mathcal{N}$, $q^{\mu}$, $\nu^{\mu}$, and $\tau^{\mu\nu}$ in terms of the sources. These are the constitutive relations. To proceed we compute derivatives of $T$, $\mu^A$, and the fluid velocity. We find
\begin{subequations}
\label{E:Dx}
\begin{align}
\partial_{\mu}T &= -T a_{\mu}, &\partial_{\mu}\mu^A &= -\mu^A a_{\mu}+E^A_{\mu},   \\
\nabla_{\mu}u_{\nu}&=-u_{\mu}a_{\nu}+\omega_{\mu\nu},&E^A_{\mu}&=F^A_{\mu\nu}u^{\nu},
\end{align}
\end{subequations}
are identically satisfied with
\begin{align}
a^{\mu}=u^{\nu}\nabla_{\nu}u^{\mu}, \,\,\,\, \omega^{\mu\nu}=\frac{\Delta^{\mu\rho}\Delta^{\nu\sigma}}{2}(\nabla_{\rho}u_{\sigma}-\nabla_{\sigma}u_{\rho}).
\end{align}
The independent tensors with one derivative are listed in Table~\ref{T:FOtensors}. 
The pseudovectors are $\tilde{v}_1^{\mu,A}=B^{\mu,A}$, the magnetic field, and $\tilde{v}_2^{\mu}=w^{\mu}$ the vorticity of the fluid. To $O(\partial)$ we then write
\begin{align}
\begin{split}
\label{E:constitutive}
\mathcal{E} & = \epsilon, \qquad \mathcal{P} = P, \qquad \mathcal{N} = \rho,\qquad \tau^{\mu\nu}=0, \\
q^{\mu} &= \gamma_i v_i^{\mu}+\tilde{\gamma}_i \tilde{v}_i^{\mu}, \qquad \nu^{\mu,A}  = \delta^A_i v_i^{\mu} + \tilde{\delta}^A_i \tilde{v}_i^{\mu},
\end{split}
\end{align}
where the $\gamma$'s, $\tilde{\gamma}$'s, $\delta$'s, and $\tilde{\delta}$'s are functions of $T$ and $\mu^A$.

\begin{table}
\begin{tabular}{|r|c|c|}
\hline
 & $1$ & $2$ \\
\hline
vectors $(v_i^{\mu})$ & $E^A_{\mu}$ & $a_{\mu}$ \\
\hline
psuedovectors $(\tilde{v}_i^{\mu})$ & $\frac{1}{2\sqrt{-g}}\epsilon^{\mu\nu\rho\sigma}u_{\nu}F^A_{\rho\sigma}$ & $\frac{1}{\sqrt{-g}}\epsilon^{\mu\nu\rho\sigma}u_{\nu}\partial_{\rho}u_{\sigma}$ \\
\hline
\end{tabular}
\caption{\label{T:FOtensors}The independent first-order tensors.}
\end{table}

We continue by treating the Ward identities~\eqref{E:Jcons} and~\eqref{E:Tcons} as equations of motion. Ordinarily, we solve for the hydrodynamic variables in the presence of external fields. In this instance, the conservation equations leads to differential equations for the coefficients in the constitutive relations. We note that the gradients of the first-order tensors satisfy some simple relations
\begin{align}
\begin{split}
\label{E:useful}
\nabla_{\mu}a^{\mu} & = u^{\mu}u^{\nu}R_{\mu\nu} + \omega^{\mu\nu}\omega_{\nu\mu}, \\
u^{\nu}\nabla_{\nu}a^{\mu}& = u^{\mu}a^2-\omega^{\mu\nu}a_{\nu}, \\
 u^{\nu}\nabla_{\nu}E^{\mu,A}&=u^{\mu}a^{\nu}E^A_{\nu}-\omega^{\mu\nu}E^A_{\nu}, \\
\nabla_{\mu}B^{\mu,A} & = B^{A}_{\mu}a^{\mu}-E^A_{\mu}w^{\mu},  \\
u^{\nu}\nabla_{\nu}B^{\mu,A} &= u^{\mu}B^A_{\nu}a^{\nu}-\omega^{\mu\nu}B^A_{\nu},\\
\nabla_{\mu}w^{\mu} &= 2 a_{\mu}w^{\mu}, \qquad u^{\nu}\nabla_{\nu}w^{\mu} = u^{\mu}a_{\nu}w^{\nu}, \\
\epsilon^{\mu\nu\rho\sigma}u_{\nu}B_{\rho}^Aw_{\sigma}&=2\omega^{\mu\nu}B_{\nu}^A, \,\,\,\, \epsilon^{\mu\nu\rho\sigma}u_{\nu}w_{\rho}\omega_{\sigma\tau}=0,
\end{split}
\end{align}
where $R_{\mu\nu}$ is the Ricci tensor. Applying~\eqref{E:Dx} and~\eqref{E:useful} to the conservation equations leads to a sum of coefficients, each of which multiplies an independent second-order tensor. Each such coefficient must vanish, which leads to a number of equations,
\begin{align}
\begin{split}
\label{E:consEqns}
\gamma_1^A=\gamma_2=\delta_1^{AB}=\delta_2^A &= 0, \\
 \frac{\partial\tilde{\delta}_1^{AB}}{\partial\mu^C}-C^{ABC}=\frac{\partial\tilde{\delta}_2^A}{\partial\mu^B}-\tilde{\delta}_1^{AB}&=0, \\
\frac{\partial\tilde{\gamma}_1^A}{\partial\mu^B}-\tilde{\delta}_1^{AB}=\frac{\partial\tilde{\gamma}_2}{\partial\mu^A}-\tilde{\delta}_2^{A}-\tilde{\gamma}_1^A&=0, \\
T\frac{\partial\tilde{\delta}^{AB}_1}{\partial T}+\mu^C\frac{\partial\tilde{\delta}_1^{AB}}{\partial\mu^C}-\tilde{\delta}_1^{AB}&=0, \\
T\frac{\partial\tilde{\delta}^{A}_2}{\partial T}+\mu^B\frac{\partial\tilde{\delta}_2^{A}}{\partial\mu^B}-2\tilde{\delta}_2^{A}&=0, \\
T\frac{\partial\tilde{\gamma}^{A}_1}{\partial T}+\mu^B\frac{\partial\tilde{\gamma}_1^{A}}{\partial\mu^B}-2\tilde{\gamma}_1^{A}&=0, \\
T\frac{\partial\tilde{\gamma}_2}{\partial T}+\mu^A\frac{\partial\tilde{\gamma}_2}{\partial\mu^A}-3\tilde{\gamma}_2&=0.
\end{split}
\end{align}
When~\eqref{E:consEqns} holds, the Ward identities~\eqref{E:Jcons} and~\eqref{E:Tcons} are exactly solved for the first-order constitutive relations~\eqref{E:constitutive}.

We compute the $ra..a$ functions by varying one-point functions (which we view as expectation values of $r$ operators) with respect to external fields~\cite{Moore:2010bu} (which we view as $a$-type sources). Those variations are easy to perform in this frame. We obtain zero-frequency $n$-point fuctions by directly varying~\eqref{E:TJdecomp} and~\eqref{E:constitutive}. For instance, we find the following two-point functions to $O(k)$,
\begin{align}
\begin{split}
G_R^{iA,jB}(\k) &= -i \epsilon^{ijk}k_k \tilde{\delta}_1^{AB}, \\
G_R^{iA,0j}(\k) &= -i \epsilon^{ijk}k_k \tilde{\delta}_2^A,\\
G_R^{0i,jA}(\k) &= - i \epsilon^{ijk}k_k \tilde{\gamma}_1^A, \\
G_R^{0i,0j}(\k) &= P\delta^{ij}- i \epsilon^{ijk}k_k \tilde{\gamma}_2.
\end{split}
\end{align}
By~\eqref{E:recip} we then have $\delta_2^A=\gamma_1^A$. Matching to~\eqref{E:G2pt} then gives
\begin{equation}
\sigma_1^{AB}=\tilde{\delta}_1^{AB}, \qquad \sigma_2^A = \tilde{\delta}_2^A, \qquad \sigma_3 = \tilde{\gamma}_2,
\end{equation}
so that the constraints~\eqref{E:consEqns} precisely reproduce those~\eqref{E:eqsSigma} and~\eqref{E:eqsE} for the 
$\sigma$'s. 


\emph{Discussion.}---%
In this letter we have used Ward identities to constrain zero-frequency correlation functions. For a $3+1$-dimensional theory with $U(1)^3$ anomalies, the $O(k)$ parts of three-point functions of the stress tensor and currents are determined by conservation in terms of the anomaly coefficients and $O(k)$ parts of two-point functions.~\eqref{E:TmuP} leads to differential equations that relate two-point functions to three-point functions. As a result the $O(k)$ terms in two and three-point functions of $\T$ and $\J$ are uniquely fixed up to integration constants. The result matches the hydrodynamic~\cite{Son:2009tf,Neiman:2010zi} and holographic~\cite{Amado:2011zx} calculation up to additional CPT-violating integration constants $f_1^{AB}$~\eqref{E:sigmas}.


Zero-frequency correlation functions encode the thermodynamic response of a fluid. In this instance, a fluid may be subjected to a magnetic field $B^{\mu}$ or vorticity $w^{\mu}$ and remain in equilibrium. When parity is not a symmetry, there may be charge and momentum currents directed along $B^{\mu}$ and $w^{\mu}$. This calculation is telling us that the $O(B)$ and $O(w)$ response of those currents is fixed by a consistent description of thermodynamics in the presence of background fields.

One significant question remains: what is the role of the CPT-preserving integration constant $f_2^A$~\eqref{E:sigmas}? At weak~\cite{Landsteiner:2011cp} and strong~\cite{Landsteiner:2011iq} coupling, it has been found to be proportional to the mixed anomaly coefficient. Is that result fixed by Ward identities? The hydrodynamic calculation sheds some light on this question. By exactly solving the conservation equations to $O(\partial^2)$ and imposing~\eqref{E:TmuP} at the outset, we have implicitly enforced the Ward identities for gauge/diffeomorphism invariance on the $O(k)$ parts of all zero-frequency $n$-point functions of $\T$ and $\J$. At the end of the day, $f_2^A$ is left unfixed by the zero-frequency conditions~\eqref{E:recip} and~\eqref{E:TmuP}. Three logical possibilities remain: (i.) $f_2^A$ is fixed by another zero-frequency condition, (ii.) $f_2^A$ is fixed by a finite-frequency property like the Onsager relations, or (iii.) $f_2^A$ is unfixed.

\emph{Acknowledgments.}---%
It is a pleasure to thank M.~Kaminski, Z.~Komargodski, P.~Kovtun, K.~Landsteiner, R.~Meyer, A.~Ritz, D.~Son, and especially A.~Yarom for stimulating discussions. This work was initiated at the Perimeter Institute for Theoretical Physics and was supported in part by NSERC, Canada.
\bibliography{anomalous_refs}

\end{document}